\documentclass[a4paper,12pt]{article}

\usepackage{graphicx}
\usepackage[namelimits]{amsmath}
\usepackage{amssymb}
\usepackage{amsthm}
\usepackage{fancybox}
\usepackage{psfrag}

\newcommand{\sn}{\mathrm{sn}}
\newcommand{\cn}{\mathrm{cn}}

\newcommand{\sech}{\mathrm{sech}}
\newcommand{\vb}[1]{{\mbox{\boldmath$#1$}}}

\newtheorem*{conjecture}{Conjecture}

\begin{document}

\title{Stationary states of the Gross-Pitaevskii equation with linear 
counterpart}
\author{
Roberto D'Agosta$^a$, Boris A. Malomed$^{b}$, Carlo Presilla$^{a,c}$ \\
$^a$Dipartimento di Fisica, Universit\`a di Roma ``La Sapienza'',\\
Piazzale A. Moro 2, Roma 00185, Italy \\
$^b$Department of Interdisciplinary Studies, Faculty of Engineering,\\ 
Tel Aviv University, Tel Aviv 69978, Israel \\
$^c$Istituto Nazionale per la Fisica della Materia, Unit\`a di Roma I }
\date{ }%July 4, 2000}
\maketitle

\begin{abstract}
We study the stationary solutions of the Gross-Pitaevskii equation
that reduce, in the limit of vanishing non-linearity, to the 
eigenfunctions of the associated Schr\"odinger equation.
By providing analytical and numerical support, we conjecture an existence 
condition for these solutions in terms of the ratio between their proper 
frequency (chemical potential) and the corresponding linear eigenvalue. 
We also give approximate expressions for the stationary solutions which
become exact in the opposite limit of strong non-linearity.
For one-dimensional systems these solutions have the form of a chain
of dark or bright solitons depending on the sign of the non-linearity.
We demonstrate that in the case of negative non-linearity 
(attractive interaction) the norm of the solutions is always bounded
for dimensions greater than one.
\end{abstract}

\leftline{PACS: 03.65.Ge, 03.75.Fi, 47.20.Ky}
\vspace{0.5cm}

\section{Introduction}

Recent achievement of Bose-Einstein condensation (BEC) in gases of 
alkali atoms has generated an impressive amount of experimental and
theoretical works \cite{isw,stringa}. 
In these systems the condensate is usually described by the so-called 
Gross-Pitaevskii equation (GPE), a Schr\"odinger equation with a local 
cubic non-linear term which represents the interaction among the bosons
in a mean field approximation. 
GPE effectively reproduces the ground state properties of a condensed 
boson gas confined by an external potential at zero temperature 
\cite{stringa}.
In the framework of linear response theory, the mean field approximation
also allows to evaluate the spectrum of the excitations in presence of 
an external time-dependent perturbation \cite{ruprec,jin}.
On the other hand, GPE also appears in the description of other physical 
systems, like nonlinear optics \cite{kld}, 
molecular physics \cite{els}, {\it etc.}.

In this paper we study general properties of the stationary 
solutions of GPE both in the case of repulsive and attractive interaction.
Besides its mathematical interest, this study is relevant in the search 
of the so-called vortex states and, in general, in understanding the 
dynamical properties of condensates. 
We choose to work in the grand-canonical ensemble, 
that is we fix the chemical potential $\mu$ of the system, 
{\it i.e.} the proper frequency for the time evolution, and derive the 
number of particles corresponding to each stationary solution. 
In particular, we study the stationary solutions of GPE that have a 
linear counterpart, in the sense that they reduce to the eigenstates of 
the linear Schr\"odinger equation which is the limit of GPE for 
vanishing interaction. 
For these states we conjecture an existence condition which depend on the
ratio between the chemical potential and the corresponding
eigenvalue of the associated Schr\"odinger equation.
We give a proof of this conjecture for the node-less
state of a system in presence of a general external potential and verify 
it for the exactly solvable case of a one-dimensional, infinitely deep, 
square well.  
We also provide numerical evidence of the validity of the conjecture by 
studying systems with harmonic potentials in different dimensions.    

As a consequence of the above conjecture, we find that in the case of 
attractive condensates there exists a range of the chemical potential 
$\mu$ in which the node-less stationary solution does not exist and the
lowest-energy state has one or more nodes. This may be relevant for the 
observation of stable vortex states.

We also study the limit of strong non-linearity of GPE obtained for 
large values of the modulus of the chemical potential $\mu$.
In this limit a Thomas-Fermi approximation holds for repulsive systems,
while for attractive systems the solutions become independent of the 
external potential. 
In the one-dimensional case the corresponding approximate solutions 
have the form of a chain of dark or bright solitons depending on the 
sign of the non-linearity.
We use these asymptotically exact expressions to establish 
that the number of particles in the ground state of an attractive
condensate is always bounded for dimensions greater than one
in agreement with previous numerical results \cite{bras1}.

\section{The linear limit}
We consider the Gross-Pitaevskii equation \cite{gross,pitaevskii} 
describing, in the mean field approximation, a system of interacting 
particles confined by an external potential $V(\vb{x})$
\begin{equation}
i\hbar\frac{\partial \Psi(\vb{x},t)}{\partial t}=-\frac{\hbar^2}{2m}\nabla^2\Psi(\vb{x},t)+U_0|\Psi(\vb{x},t)|^2\Psi(\vb{x},t)+V(\vb{x})\Psi(\vb{x},t),
\label{gpetime}
\end{equation}
with $\vb{x}\in\mathbb{R}^d$. 
The constant $U_0$ is positive (negative) in the case of repulsive (attractive) interaction.
Equation (\ref{gpetime}) has two conserved quantities, namely the number of particles (squared norm)
\begin{equation}
N[\Psi]=\int |\Psi(\vb{x},t)|^2 d \vb{x}
\label{norm}
\end{equation}
and the energy 
\begin{equation}
E[\Psi]=\int\left[\frac{\hbar^2}{2m}|\nabla\Psi(\vb{x},t)|^2+\frac{U_0}{2}|\Psi(\vb{x},t)|^4+V(\vb{x})|\Psi(\vb{x},t)|^2\right]d\vb{x}.
\label{energy}
\end{equation}

The stationary states of Eq. (\ref{gpetime}),  
$\Psi(\vb{x},t)=\exp \left( -\frac{i}{\hbar}\mu t \right) \psi(\vb{x})$, where $\mu$ is the chemical potential are determined by the equation
\begin{equation}
-\frac{\hbar^2}{2m}\nabla^2\psi(\vb{x})+U_0|\psi(\vb{x})|^2\psi(\vb{x})+V(\vb{x})\psi(\vb{x})-\mu\psi(\vb{x})=0,
\label{gpe}
\end{equation}
i.e. as critical points of the grand-potential functional
\begin{eqnarray}
\Omega[\psi]&=&\int\left[\frac{\hbar^2}{2m}|\nabla\psi(\vb{x})|^2+\frac{U_0}{2}|\psi(\vb{x})|^4+\left(V(\vb{x})-\mu\right)|\psi(\vb{x})|^2\right]d\vb{x}\nonumber \\
&=&E[\psi]-\mu N[\psi].
\label{functional}
\end{eqnarray}
It is simple to show \cite{jo-fa} that if $\psi$ is a solution of 
(\ref{gpe}) then 
\begin{equation}
\Omega[\psi]=-\frac{U_0}{2}\int|\psi(\vb{x})|^4d\vb{x}.
\label{funcond}
\end{equation}

We will look for the solutions of (\ref{gpe}) corresponding to a given chemical potential $\mu$. 
In this paper, we concentrate on solutions which admit a linear 
counterpart in the sense that they reduce, in a proper limit, to the 
eigenfunctions of the associated linear problem 
\begin{equation}
-\frac{\hbar^2}{2m}\nabla^2\phi_n(\vb{x})+V(\vb{x})\phi_n(\vb{x})-\mathcal{E}_n\phi_n(\vb{x})=0.
\label{linearsch}
\end{equation}
Here we suppose that $\mathcal{E}_0\leq \mathcal{E}_1\leq \ldots \leq \mathcal{E}_n$ and $\left\{\phi_n(\vb{x})\right\}$ is a hortonormal base
with $\phi_0(\vb{x})$ positive and bounded.
Solutions without linear counterpart will be discussed in another paper \cite{damopr}. 

By substituting $\psi(\vb{x})=\sqrt{N(\mu)}\chi(\vb{x})$ in (\ref{gpe}) 
with $||\chi||=1$, we have 
\begin{equation}
-\frac{\hbar^2}{2m}\nabla^2\chi(\vb{x})+U_0 N(\mu) |\chi(\vb{x})|^2\chi(\vb{x})+(V(\vb{x})-\mu)\chi(\vb{x})=0.
\label{qlinear}
\end{equation}
If the number of particles is sufficiently small, the nonlinear term in (\ref{qlinear}) can be neglected and $\chi$ approximated by $\phi_n$. 
By substituting $\chi$ with $\phi_n$, Eq. (\ref{qlinear}) provides the following relation between the chemical potential $\mu$ and the corresponding norm $N(\mu)$
\begin{equation}
\mu\simeq\mathcal{E}_n+U_0 N(\mu) ||\phi_n^2||^2.
\label{mun}
\end{equation}
Equation (\ref{mun}) suggests the following conjecture for the existence
 of solutions of (\ref{gpe}) with linear counterpart 
\begin{conjecture}
For $U_0>0$ $(U_0<0)$, solutions with linear limit $\psi\simeq \sqrt{N(\mu)}\phi_n$ 
exist only if $\mu>\mathcal{E}_n$ $(\mu<\mathcal{E}_n)$.
Moreover $N(\mu)\to 0$ for $\mu \to \mathcal{E}_n$.
\end{conjecture}\noindent In Appendix \ref{appA} we give a general proof of this conjecture in the case $n=0$.

The conjecture can be verified analytically in the case of a 1-dimensional system confined in a box of size $L$, i.e. with 
\begin{equation}
V(x)=\left\{
\begin{array}{cc}
0 & |x|<L/2\\
\infty &  |x|>L/2
\end{array}
\right. .
\end{equation}
For this problem the solutions of (\ref{gpe}) are known \cite{lcarr}. 
In the case $U_0>0$ they are given by the Jacobi elliptic functions 
\begin{equation}
\psi_n(x)=A~\sn \left(\left. 2(n+1)\mbox{K}(p)\left(\frac{x}{L}+\frac{1}{2}\right)\right|p\right),
\label{sol1d}
\end{equation}
where 
\begin{equation}
\mbox{K}(p)=\int_0^{\frac{\pi}{2}}\frac{1}{\sqrt{1-p\sin^2\theta}}d\theta
\end{equation}
is the complete elliptic integral of the first kind with modulus 
$p\in [0,1]$, and  $n=0,1,2,\ldots$. 
By substituting (\ref{sol1d}) into (\ref{gpe}), one finds the conditions
\begin{eqnarray}
A^2&=&\frac{\hbar^2}{mU_0L^2}~p~(2(n+1)\mbox{K}(p))^2,\\
\mu &=& \frac{\hbar^2}{mL^2}~\frac{p+1}{2}~(2(n+1)\mbox{K}(p))^2.
\label{mu1}
\end{eqnarray}
The number of particles and the energy are given by
\begin{equation}
N(\mu)=\frac{\hbar^2}{mU_0L}~(2(n+1)\mbox{K}(p))^2\left(1-\frac{\mbox{E}(p)}{\mbox{K}(p)}\right),
\label{numsqwep}
\end{equation}
\begin{equation}
E(\mu)=N\mathcal{E}_0~\frac{(n+1)^2}{3}
 \left(\frac{2\mbox{K}(p)}{\pi}\right)^2~ \frac{p+(p+1)\left(1-\frac{\mbox{E}(p)}{\mbox{K}(p)}\right)}{1-\frac{\mbox{E}(p)}{\mbox{K}(p)}},
\label{ensqwep}
\end{equation}
where   
\begin{equation}
\mbox{E}(p)=\int_0^{\frac{\pi}{2}}\sqrt{1-p\sin^2\theta}~d\theta
\end{equation}
is the complete elliptic integral of the second kind
with $p$ determined in terms of $\mu$ by Eq. (\ref{mu1}).
Since $\mbox{K}(p)$ increases monotonously from $\mbox{K}(0)=\pi/2$, for a given $n$ Eq. (\ref{mu1}) has solution only if 
\begin{equation}
\mu\geq \mathcal{E}_n\equiv \frac{(n+1)^2\pi^2\hbar^2}{2mL^2}
\end{equation}
which complies with the conjecture formulated above.
The same conclusion can also be reached by using the theorems of 
\cite{jo-fa}.

In the linear limit $\mu \to \mathcal{E}_n$, the solutions (\ref{sol1d}) 
reduce to the eigenfunctions of the associated Schr\"odinger equation 
\begin{equation}
\frac{1}{\sqrt{N(\mu)}}\psi_n(x)\stackrel{\mu \to \mathcal{E}_n}{\longrightarrow}\sqrt{\frac{2}{L}}\sin\left[\left(\frac{x}{L}+\frac{1}{2}\right)(n+1)\pi\right].
\label{linlim}
\end{equation}  
In the opposite limit of strong nonlinearity, $\mu \gg \mathcal{E}_n$, 
we get from (\ref{sol1d}) the dark soliton solutions 
\begin{eqnarray}
\psi_n(x)&\stackrel{\mu \gg \mathcal{E}_n}{\longrightarrow}&\sqrt{\frac{\mu}{U_0}}~\prod_{k=0}^{n+1}\tanh \left(\frac{\sqrt{m\mu}}{\hbar}(x-x_k)\right),\\
&&x_k=-\frac{L}{2}+\frac{L}{n+1}k.
\end{eqnarray}   

Similar results are obtained in the case $U_0<0$. The solutions of (\ref{gpe}) are now given by
\begin{equation}
\psi(x)=A~\cn\left(\left. 2(n+1)\mbox{K}(p)\left(\frac{x}{L}+\frac{1}{2}\right)+\mbox{K}(p)\right|p\right)
\label{sol1d_neg}
\end{equation}
with the conditions
\begin{eqnarray}
A^2&=&-\frac{\hbar^2}{mU_0L^2}~p~(2(n+1)\mbox{K}(p))^2\\
\mu&=& \frac{\hbar^2}{mL^2}~\frac{1-2p}{2}~(2(n+1)\mbox{K}(p))^2.
\label{mu2}
\end{eqnarray}
Number of particles and the energy become
\begin{equation}
N(\mu)=-\frac{\hbar^2}{mU_0L}~(2(n+1)\mbox{K}(p))^2\left(p-1+\frac{\mbox{E}(p)}{\mbox{K}(p)}\right),
\label{numsqwen}
\end{equation}
\begin{equation}
E(\mu)=N\mathcal{E}_0~\frac{(n+1)^2}{3} \left(\frac{2\mbox{K}(p)}{\pi}\right)^2~\frac{p(1-p)+(1-2p)\left(p-1+\frac{\mbox{E}(p)}{\mbox{K}(p)}\right)}{p-1+\frac{\mbox{E}(p)}{\mbox{K}(p)}},
\label{ensqwen}
\end{equation}
where $p$ is determined by Eq. (\ref{mu2}). 
Since $(1-2p)\mbox{K}(p)$ decreases monotonously for $p\in [0,1]$, the $n$-node solution exists only if $\mu\leq \mathcal{E}_n$ as conjectured above.

For $\mu \to \mathcal{E}_n$ the solutions (\ref{sol1d_neg}) 
have the same limit (\ref{linlim}). 
For $-\mu \gg \mathcal{E}_n$, we get the bright soliton solutions 
\begin{eqnarray}
\psi_n(x)&\stackrel{-\mu \gg \mathcal{E}_n}{\longrightarrow}&\sqrt{\frac{2 \mu}{U_0}}~\sum_{k=0}^{n}(-1)^k \sech \left(\frac{\sqrt{-2 m\mu}}{\hbar}(x-x_k)\right),\\
&&x_k=-\frac{L}{2}+\frac{L}{n+1}\left(k+\frac{1}{2}\right).
\end{eqnarray}

\begin{figure}[ht!]
\begin{center}
\psfrag{assex}[][]{$\mu/\mathcal{E}_0$}
\psfrag{assey}[][]{$N(\mu)~m|U_0|L/\hbar^2$}
\includegraphics[width=8.5cm]{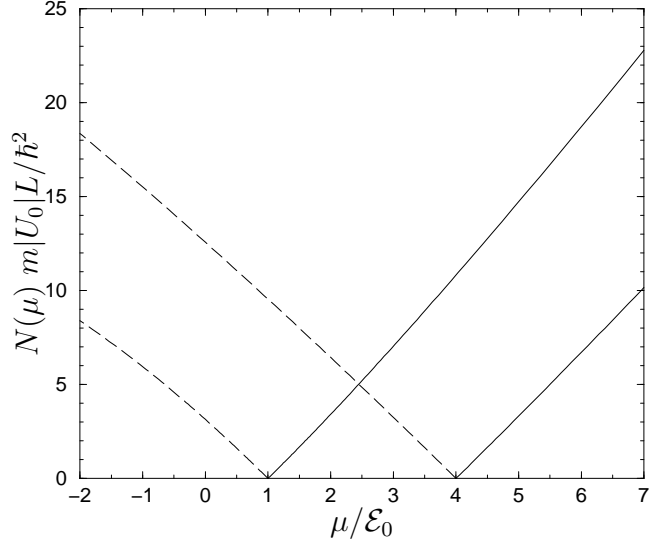}
\caption{Number of particles $N$ as a function of the chemical potential $\mu$ for the one dimensional square well. The solid and dashed lines are given by Eqs. (\protect{\ref{numsqwep}}) and (\protect{\ref{numsqwen}}), respectively. The two curves correspond to the states $n=0$ and $n=1$.}
\label{figure1}
\end{center}
\end{figure}

\begin{figure}[ht!]
\begin{center}
\psfrag{assex}[][]{$\mu/\mathcal{E}_0$}
\psfrag{assey}[][]{$E(\mu)/N(\mu)\mathcal{E}_0$}
\includegraphics[width=8.5cm]{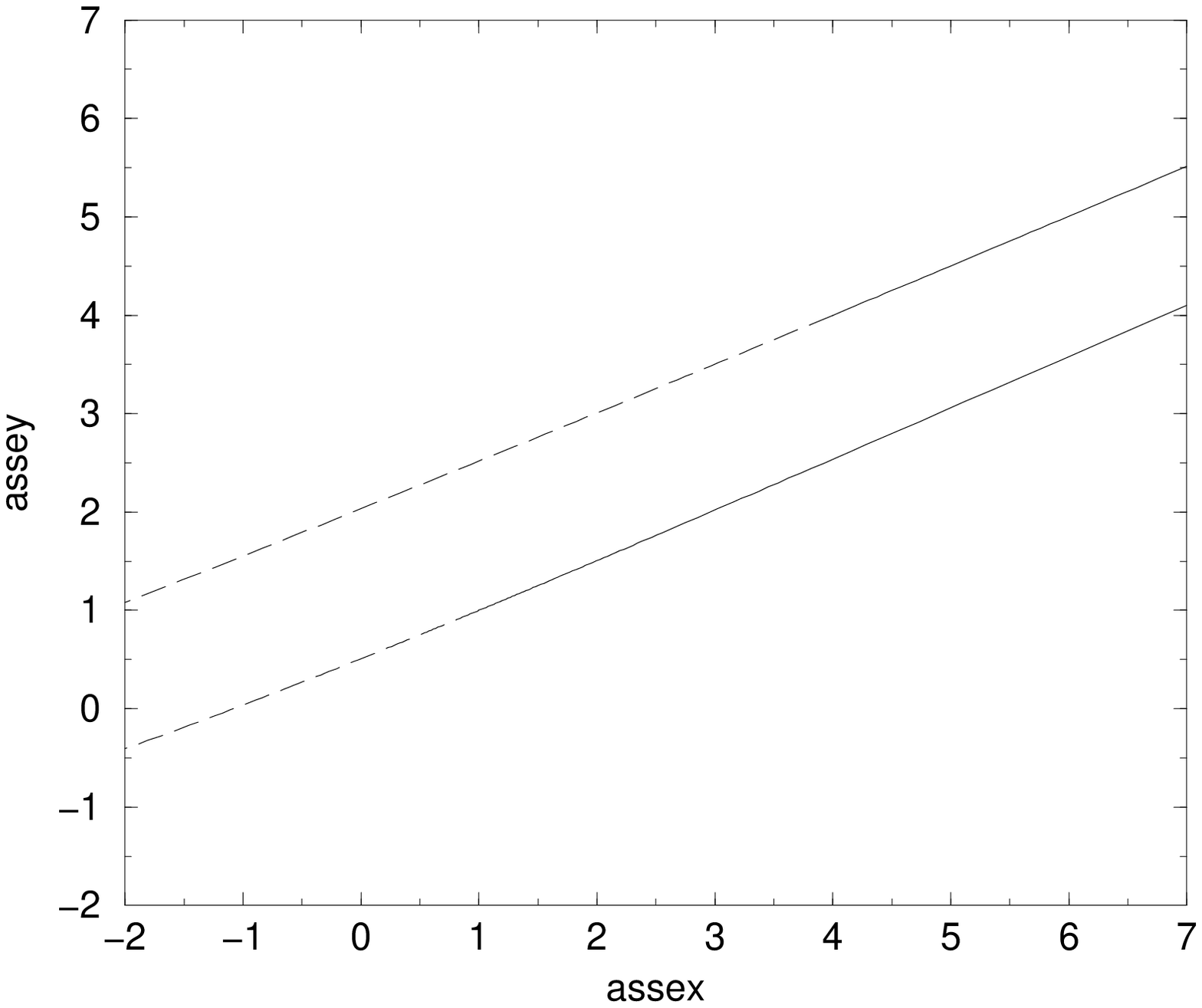}
\caption{Single-particle energy $E/N$ as a function of the chemical potential $\mu$ for the same states of Fig. \protect{\ref{figure1}}. The solid and dashed lines are given by Eqs. (\protect{\ref{ensqwep}}) and (\protect{\ref{ensqwen}}), respectively.}
\label{figure4}
\end{center}
\end{figure}

In Fig. \ref{figure1} we show the behaviour of $N(\mu)$ evaluated according to (\protect{\ref{numsqwep}}) and (\protect{\ref{numsqwen}}) 
for the states $n=0$ and $n=1$. 
Note that $N(\mu) \to 0$ for $\mu \to \mathcal{E}_n$. The single-particle energy, $E(\mu)/N(\mu)$, for the same states is shown in Fig. \ref{figure4}

We have verified the above conjecture with numerical and analytical methods also in the case of a quadratic potential $V(\vb{x})=\frac{1}{2}m\omega^2 \sum_{i=1}^d x_i^2$, with $d=1,2,3$. For example, in the case $d=1$ consider the following 
Ansatz for the solutions of (\ref{gpe})
\begin{equation}
\psi_n(x)=a_n~\exp\left(-\frac{x^2}{2 b_n^2}\right)H_n\left(\frac{x}{b_n}\right),
\end{equation}
where $H_n\left(x\right)$ is the Hermite polynomial of degree $n$ 
and $a_n$, $b_n$ are real constants. 
Extremization of the functional $\Omega$ with respect to $a_n$ and $b_n$ leads to
\begin{eqnarray}
a_n^2 &=&\frac{\mu}{U_0}~\frac{2^{n+1}n!\left(8-\sqrt{4+15(2n+1)^2\eta^2}\right)}{15g_n},\\
b_n^2 &=&\frac{\hbar^2}{m\mu}~\frac{2+\sqrt{4+15(2n+1)^2\eta^2}}{5(2n+1)\eta},
\end{eqnarray}
where $\eta=\hbar^2\omega^2/\mu^2$ and $g_n=\int_{-\infty}^{\infty}H_n(x)^4dx$.
If $U_0>0$, the condition $a_n^2>0$ implies 
$\mu>\left(n+\frac{1}{2}\right)\hbar\omega$.
If $U_0<0$, the same condition leads to 
$\mu<\left(n+\frac{1}{2}\right)\hbar\omega$.
Note that in the linear limit  $\mu\rightarrow (n+\frac{1}{2})\hbar\omega$, we have $N\propto a_n^2 \to 0$ and $b_n^2\rightarrow \hbar/m\omega$.

Analogously, in the case $d=2$ consider the Ansatz
\begin{equation}
\psi_n(x_1,x_2)=a_n~\exp\left(-\frac{x^2}{2b_n^2}\right)F\left(-n,|m|+1,\left(\frac{r}{b_n}\right)\right)\left(\frac{r}{b_n}\right)^{|m|}e^{im\theta},
\end{equation}
where $r^2=x_1^2+x_2^2$, $\tan \theta=x_2/x_1$ and $F(n,m,r)$ is the confluent hypergeometric function \cite{abramowitz}. 
The condition $a_n^2>0$ is now equivalent to $\mu>\mathcal{E}_{n,m}$ if $U_0>0$, and $\mu<\mathcal{E}_{n,m}$ if $U_0<0$, where $\mathcal{E}_{n,m}=\left(2n+|m|+1\right)\hbar\omega$ are the eigenvalues of the associated Schr\"odinger equation.

\begin{figure}[ht!]
\begin{center}
\psfrag{assex}[][]{$\mu/\hbar\omega$}
\psfrag{assey}[][]{$N(\mu)~m|U_0|/2\pi\hbar^2$}
\includegraphics[width=8.5cm]{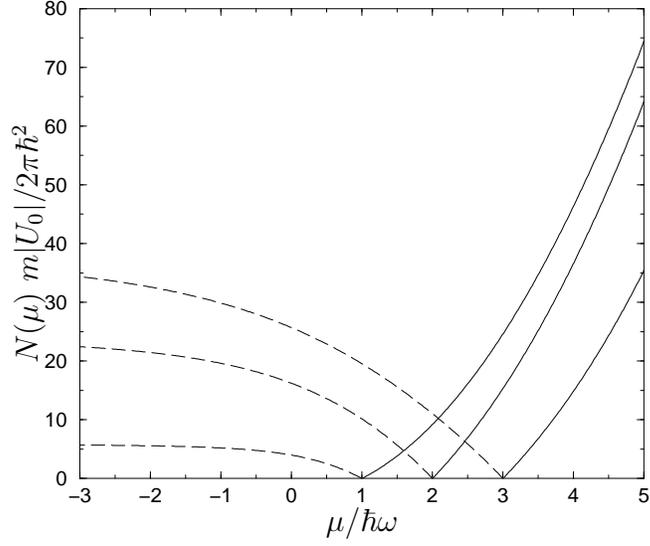}
\caption{Number of particles $N$ as a function of the chemical potential 
$\mu$ for a two dimensional quadratic potential. Solid and dashed lines
are obtained by the solving numerically Eq. (\protect{\ref{gpe}}) for 
$U_0>0$ and $U_0<0$, respectively. The three curves correspond to the states $(n,m)=(0,0)$, $(0,1)$ and $(1,0)$.}
\label{figure2}
\end{center}
\end{figure}

\begin{figure}[ht!]
\begin{center}
\psfrag{assex}[][]{$\mu/\hbar\omega$}
\psfrag{assey}[][]{$E(\mu)/N(\mu)\hbar\omega$}
\includegraphics[width=8.5cm]{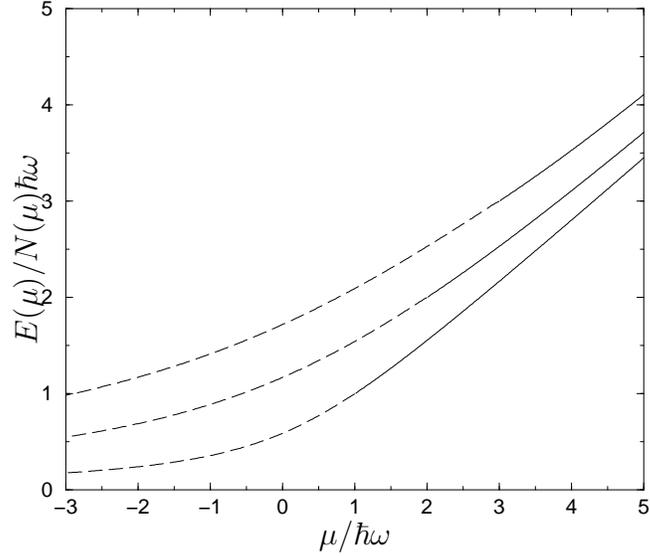}
\caption{Single-particle energy $E/N$ as a function of the chemical potential 
$\mu$ for the same states of Fig. \protect{\ref{figure2}}.}
\label{figure5}
\end{center}
\end{figure}

Equation (\ref{gpe}) has also been solved numerically with a standard relaxation algorithm \cite{numrec}. In Fig. \ref{figure2} we show the number of particles obtained as a function of the chemical potential $\mu$ for the states $(n,m)=(0,0)$, $(0,1)$ and $(1,0)$ in the case of a two-dimensional quadratic potential. The single-particle energy for the same states is shown in Fig. \ref{figure5}. Similar results are obtained for $d=1$ and $d=3$.

Figures \ref{figure1}-\ref{figure5} allow us to emphasize a possibly 
important consequence of the above conjecture. 
In the case $U_0<0$, the node-less solution exists only for 
$\mu<\mathcal{E}_0$. 
Therefore, in the range $\mathcal{E}_0<\mu<\mathcal{E}_1$ the state with minimal energy is $\Psi_1$. This implies that controlling the chemical potential it is possible to obtain a condensate with a node or a vortex in the ground state.    

\section{The strongly non linear limit}
The conjecture discussed so far concerned the behaviour of the solutions of Eq. (\ref{gpe}) in the linear limit.
An approximate expression of the solutions of (\ref{gpe}) is possible also
in the opposite limit $|\mu| \to \infty$.
Let us consider first the case $U_0>0$. The repulsive interaction tends to delocalize the solutions so that the Thomas-Fermi approximation holds \cite{huang,baym}. In this case the gradient term in Eq. (\ref{gpe}) can be neglected and $\psi$ is determined by
\begin{equation}
U_0|\psi(\vb{x})|^2\psi(\vb{x})+V(\vb{x})\psi(\vb{x})-\mu\psi(\vb{x})=0.
\end{equation}
Therefore the ground state solution can be approximated as
\begin{equation}
\psi_0(\vb{x})=\left\{
\begin{array}{ll}
\sqrt{\left(\mu-V(\vb{x})\right)/U_0}& \mu>V(\vb{x})\\
0 & \mu<V(\vb{x})
\end{array}  
\right. .
\label{tf0}
\end{equation}
In the one-dimensional case, $n$-node solutions may be approximated by a chain of dark solitons 
\begin{equation}
\psi_n(x)=\psi_0(x)\prod_{k=1}^{n}\tanh \left(\frac{\sqrt{m\mu}}{\hbar}(x-x_k)\right)
\end{equation}
with $x_k$ to be determined, for instance by extremizing the functional $\Omega$.

In the case of a quadratic potential the number of particles and the energy for the state (\ref{tf0}) are
\begin{equation}
N(\mu)=\frac{2^{\frac{d+2}{2}}}{d(d+2)}\Lambda(d)~\frac{\mu^{\frac{d+2}{2}}}{m^{\frac{d}{2}}U_0 \omega^d},
\label{nup.app}
\end{equation}
\begin{equation}
E(\mu)=N\mu~ \left(1-\frac{2}{d+4}\right),
\label{enhopo}
\end{equation}
where $\Lambda(d)$ is the volume of the unitary $d$-dimensional sphere. From Eq. (\ref{nup.app}) we see that $N$ diverges for $\mu \to \infty$. Similar results are obtained for other potentials.

In the attractive case $U_0<0$, the solutions of (\ref{gpe}) tend to localize and the Thomas-Fermi approximation fails \cite{huang}.
In this case, however, for $\mu\to -\infty$ the potential term $V\psi$ becomes negligible and Eq. (\ref{gpe}) can be approximated as
\begin{equation}
-\frac{\hbar^2}{2m}\nabla^2\psi(\vb{x})+U_0|\psi(\vb{x})|^2\psi(\vb{x})-\mu\psi(\vb{x})=0.
\label{bigmu}
\end{equation}

Recently numerical evidence has been provided that the number of particles confined in a two dimensional harmonic potential is limited in the case of attractive interaction \cite{bras1}.
This fact can be analytically understood from (\ref{bigmu}).     
With the change 
\begin{eqnarray}
\vb{x} &=& \frac{\hbar}{\sqrt{-m\mu}}~\vb{\xi}\\
\psi(\vb{x}) &=& \sqrt{\frac{\mu}{U_0}}~\phi(\vb{\xi}),
\end{eqnarray}
Eq. (\ref{bigmu}) can be rewritten in the adimensional form
\begin{equation}
-\frac{1}{2}\nabla_{\vb{\xi}}^2\phi(\vb{\xi})-|\phi(\vb{\xi})|^2\phi(\vb{\xi})+\phi(\vb{\xi})=0.
\label{eqad}
\end{equation}
Note that in the one-dimensional case, 
$n$-node solutions may be approximated by a chain of bright solitons 
\begin{equation}
\psi_n(x)=\sqrt{\frac{2 \mu}{U_0}}~\sum_{k=0}^{n}(-1)^k \sech \left(\frac{\sqrt{-2 m\mu}}{\hbar}(x-x_k)\right)
\end{equation}
with $x_k$ to be determined, for instance by extremizing the functional $\Omega$.
The number of particles corresponding to a solution of (\ref{eqad}) 
is given by
\begin{eqnarray}
N(\mu)=\int|\psi(\vb{x})|^2d\vb{x}&=&\frac{\mu}{U_0} \frac{\hbar^d}{\left(-m\mu\right)^\frac{d}{2}}\int|\phi(\vb{\xi})|^2d\vb{\xi} \nonumber\\
&=&\frac{\hbar^d}{m^{\frac{d}{2}}|U_0|}|\mu|^{\frac{2-d}{2}}\Gamma_2(d),
\label{num.app}
\end{eqnarray}
where $\Gamma_k(d)=\int|\phi(\vb{\xi})|^k~d\vb{\xi}$ is a numerical constant.
Therefore we have
\begin{equation}
\lim_{\mu\rightarrow -\infty}N(\mu)=\left\{
\begin{array}{ll}
\infty & d=1\\
\frac{\hbar^2}{m |U_0|}\Gamma_2(2) & d=2\\
0 & d\geq 3
\end{array}
\right. .
\end{equation}

\begin{figure}[ht!]
\begin{center}
\psfrag{assey}[][]{$N(\mu)$}
\psfrag{assex}[][]{$\mu/\hbar\omega$}
\includegraphics[width=8.5cm]{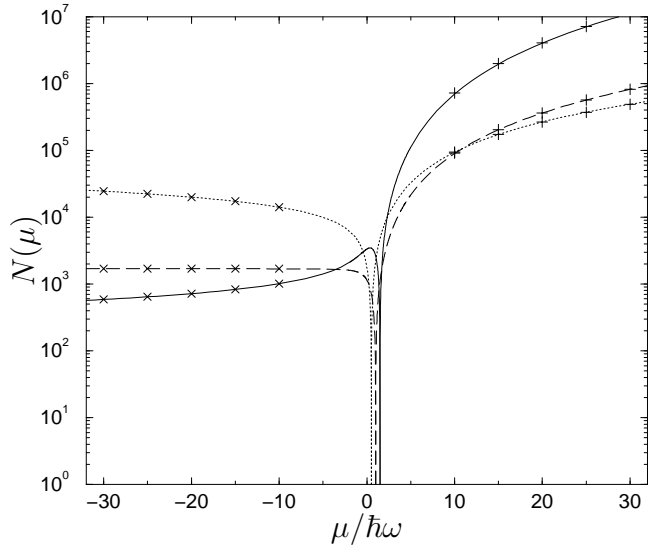}
\caption{Number of particles $N$ as a function of the chemical potential $\mu$ in the ground state of a quadratic potential with $d=1$ (dotted line), $d=2$ (dashed line) and $d=3$ (solid line). The dots $+$ and $\times$ are the analytical results (\protect{\ref{nup.app}}) and (\protect{\ref{num.app}}), respectively.}
\label{figure3}
\end{center}
\end{figure}

In Fig. \ref{figure3} we show the behaviour of $N(\mu)$ in the ground state obtained by solving numerically Eq. (\ref{gpe}) with a harmonic potential in the cases $d=1,2,3$. We have chosen the following realistic values for the parameters: $m=3.818\times 10^{26}~{\rm Kg}$, $\omega=10.0~ {\rm Hz}$ and, for $d=3$, $U_0=4\pi \hbar^2 a_s/m$ with $a_s=2.75 \times 10^{-9}~{\rm m}$ \cite{ketter}. For $d=2$ and $d=1$ we set $U_0=4\pi \hbar^2 a_s/mL$ and  $U_0=4\pi \hbar^2 a_s/mL^2$ with $L=10^{-5}~{\rm m}$ and $L^2=9 \times 10^{-10}~{\rm m^2}$. The numerical results compare very well with the analytical approximations (\ref{nup.app}) for $U_0>0$ and (\ref{num.app}) for $U_0<0$, respectively. 
In the case of Eq. (\ref{num.app}), $\Gamma_2(d)$ has been evaluated numerically. We have $\Gamma_2(1)=2.82842$, $\Gamma_2(2)=5.85044$ and $\Gamma_2(3)=6.68118$. 
Note that for $d=3$, $N(\mu)$ has a maximum and vanishes for both $\mu \to -\infty$ and $\mu \to 3/2\hbar\omega$. This implies that the function $\mu(N)$ is not single-valued but has two branches in agreement with \cite{bras2}. 

\begin{figure}[ht!]
\begin{center}
\psfrag{assey}[][]{$E(\mu)/N(\mu)\hbar\omega$}
\psfrag{assex}[][]{$\mu/\hbar\omega$}
\includegraphics[width=8.5cm]{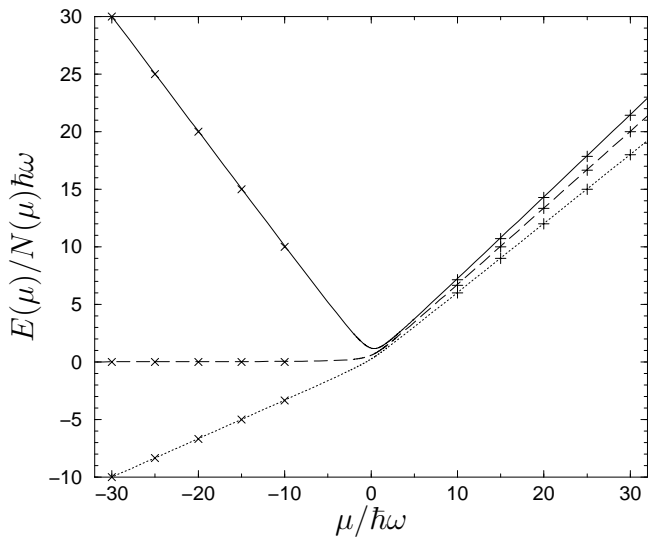}
\caption{Single-particle energy $E/N$ as a function of $\mu$ in the ground state of a quadratic potential with $d=1$ (dotted line), $d=2$ (dashed line) and $d=3$ (solid line). The dots $+$ and $\times$ are the analytical results (\protect{\ref{enhopo}}) and (\protect{\ref{enhone}}), respectively.}
\label{figure6}
\end{center}
\end{figure}

In Fig. \ref{figure6} we show the single-particle energy evaluated numerically in the same cases of Fig. \ref{figure3}. For $\mu\to \infty$ the energy diverges for any value of $d$ according to the limiting expression (\ref{enhopo}). For $\mu\to -\infty$ the behaviour of $E(\mu)$ is well described by  
\begin{equation}
E(\mu)=N\mu~ \left(1-\frac{1}{2}\frac{\Gamma_4(d)}{\Gamma_2(d)}\right)
\label{enhone}
\end{equation}
which easily stems from Eq. (\ref{eqad}).
For $d=2$ we have $\Gamma_4=2 \Gamma_2$ and hence $E(\mu)$ vanishes for $\mu\to -\infty$.

\appendix
\section{Existence of a node-less state}\label{appA}
In the following we suppose that the external potential is bounded 
from below and, for simplicity, we take $V(\vb{x})\geq 0$. We will prove that, 
with $U_0>0$, a solution of (\ref{gpe}) exists if and only if $\mu>\mathcal{E}_0$. 
The proof of the necessary condition is based on the property (\ref{funcond}). 
Let us define the functional 
\begin{equation}
Q_0[\psi]\equiv\int\left[\frac{\hbar^2}{2m}|\nabla \psi(\vb{x})|^2+(V(\vb{x})-\mu)|\psi(\vb{x})|^2\right] d \vb{x}.
\end{equation} 
We have $\Omega[\psi]=Q_0[\psi]+\frac{1}{2}U_0\int|\psi(\vb{x})|^4d \vb{x}$. 
If $Q_0[\psi]>0$, then $\psi(\vb{x})$ cannot be a solution of (\ref{gpe}).
The linear problem $Q'_0[\phi_n;\vb{x}]=k_n\phi_n(\vb{x})$, where
\begin{equation}
Q'_0[\phi_n;\vb{x}]\equiv\frac{\delta Q_0[\phi_n]}{\delta \phi_n(\vb{x})^*}=-\frac{\hbar}{2m}\nabla^2\phi_n(\vb{x})+\left(V(\vb{x})-\mu\right)\phi_n(\vb{x}),
\end{equation}
has the same eigenfunctions of (\ref{linearsch}) and the eigenvalues are  
$k_n=\mathcal{E}_n-\mu$. By decomposing a generic $\psi(\vb{x})$ as $\psi(\vb{x})=\sum_{n=0}^{\infty}c_n\phi_n(\vb{x})$, we obtain
\begin{eqnarray}
Q_0[\psi]&=&(Q'_0[\psi;\vb{x}],\psi(\vb{x})) \nonumber\\
&=&\left(\sum_{n=0}^{\infty}c_n k_n \phi_n(\vb{x}),\sum_{m=0}^{\infty}c_m \phi_m(\vb{x})\right)\nonumber \\
&\geq& k_0\sum_{n=0}^{\infty}|c_n|^2.
\end{eqnarray}
Therefore, if $\mu<\mathcal{E}_0$ we have $k_0>0$ and $Q_0[\psi]>0$.

The sufficient condition can be proved with the help of general theorems 
on elliptic differential equations 
\cite{pao}. First we look for upper and lower solutions of (\ref{gpe}). 
An upper solution $\psi_u(\vb{x})$ is defined by
\begin{equation}
-\frac{\hbar^2}{2m}\nabla^2\psi_u(\vb{x})+U_0|\psi_u(\vb{x})|^2\psi_u(\vb{x})+(V(\vb{x})-\mu)\psi_u(\vb{x})\geq 0.
\end{equation} 
For a lower solution $\psi_l(\vb{x})$ the inequality is reversed.
If a couple of ordered upper and lower solutions exist,  
{\it i.e.} $\psi_u>\psi_l$, then the existence of, at least, one solution $\psi(\vb{x})$ with $\psi_l\leq\psi\leq\psi_u$ is guaranteed \cite{pao}. 
It is simple to check that an upper solution is $\psi_u(\vb{x})=\sqrt{\mu/U_0}$. As a lower solution we choose $\psi_l(\vb{x})=\epsilon\phi_0(\vb{x})$ with
\begin{equation}
\epsilon<\min\left(\sqrt{\frac{\mu-\mathcal{E}_0}{U_0\max_\vb{x}|\phi_0(\vb{x})|^2}},\frac{\sqrt{\mu}}{\sqrt{U_0}\max_\vb{x}|\phi_0(\vb{x})|}\right)
\end{equation}
which ensures that $\psi_l<\psi_u$.

In the case  $U_0<0$, it is possible to prove that a positive solution 
of (\ref{gpe}) does not exist if $\mu>\mathcal{E}_0$.
Multiplying (\ref{gpe}) by $\phi_0(\vb{x})$ and integrating, we have
\begin{eqnarray}
0&=&\int\phi_0(\vb{x})\left[-\frac{\hbar^2}{2m}\nabla^2+U_0|\psi(\vb{x})|^2+(V(\vb{x})-\mu) \right]\psi(\vb{x}) ~d \vb{x}\nonumber\\
&=&\int\psi(\vb{x})\left[-\frac{\hbar^2}{2m}\nabla^2+U_0|\psi(\vb{x})|^2+(V(\vb{x})-\mu) \right]\phi_0(\vb{x})~d \vb{x}\nonumber\\
%=&\int\left[-\frac{\hbar^2}{2m}\psi(x)\nabla^2\phi_0(x)+U_0\phi_0(x)|\psi(x)|^2\psi(x)+\phi_0(x)(V(x)-\mu)\psi(x) \right] d x\\
&=&\int \psi(\vb{x})\left[U_0|\psi(\vb{x})|^2+(\mathcal{E}_0-\mu)\right]\phi_0(\vb{x})~d\vb{x}.
\end{eqnarray}
Therefore,
\begin{equation}
U_0\int \phi_0(\vb{x})|\psi(\vb{x})|^2\psi(\vb{x})d\vb{x}= (\mu-\mathcal{E}_0)\int\phi_0(\vb{x})\psi(\vb{x})d\vb{x}.
\label{integiden}
\end{equation}
If $\psi(\vb{x})$ is a positive function, both integrals in (\ref{integiden}) are positive and for $\mu>\mathcal{E}_0$ the equality is impossible.

\end{document}